 \documentclass[3p,times,onecolumn, round, compress]{elsarticle}
 
 \makeatletter
\def\ps@pprintTitle{%
  \let\@oddhead\@empty
  \let\@evenhead\@empty
  \let\@oddfoot\@empty
  \let\@evenfoot\@empty
}
\makeatother

\usepackage[english]{babel}
\usepackage{amssymb}
\usepackage{lipsum}
\usepackage[hyphens]{url}
\usepackage{hyperref}
\hypersetup{breaklinks=true}

\usepackage{amsmath}
\usepackage{graphicx}
\usepackage{siunitx}
\usepackage{tabularx} %
\usepackage{multirow} 
\usepackage{subcaption}
\usepackage{float}
\usepackage{tikz}
\usepackage{booktabs}

\journal{European Journal of Radiology ## AI}

\begin{document}

\begin{frontmatter}

\begin{titlepage}
    \centering
    \vspace*{2cm}
    
    {\LARGE \textbf{Advanced Deep Learning Techniques for Automated Segmentation of Type B Aortic Dissections}\par}
    \vspace{1cm}
    
    \textbf{Hao Xu\textsuperscript{a,*}, Ruth Lim\textsuperscript{b}, Brian E. Chapman\textsuperscript{a}}\\[0.5cm]
    
    \textsuperscript{a}Faculty of Engineering and IT, University of Melbourne, Parkville, VIC 3010, Australia\\[0.2cm]
    \textsuperscript{b}Melbourne Medical School, University of Melbourne, Parkville, VIC 3010, Australia\\[0.2cm]
    
    \textbf{corresponding author}: \texttt{hao.xu@unimelb.edu.au}
    
    \vspace{1cm}
    
    \textbf{Article Type:} Original Research\\[0.5cm]
    
    \textbf{Summary Statement:} This study demonstrates that the proposed framework of the deep learning model significantly improves the accuracy of aortic dissection segmentation in CT images, reducing analysis time and increasing diagnostic reliability.
    
    \vspace{0.5cm}
    
    \textbf{Key Points:}
    \begin{itemize}
        \item Proposed frameworks achieved Dice Coefficients of 0.92 for the true lumen, 0.89 for the false lumen, and 0.47 for false lumen thrombosis, outperforming previous methods.
        \item The automated segmentation reduced analysis time from 1.5 hours per scan to mere minutes.
        \item Improved segmentation accuracy aids in better diagnosis and treatment planning for patients with type B aortic dissection.
    \end{itemize}
    
    \vfill
    \date{\today}
    
\end{titlepage}

\begin{abstract} 
    \noindent \textbf{Purpose:} Aortic dissections are life-threatening cardiovascular conditions requiring accurate segmentation of true lumen (TL), false lumen (FL), and false lumen thrombosis (FLT) from CTA images for effective management. Manual segmentation is time-consuming and variable, necessitating automated solutions.

    \noindent \textbf{Materials and Methods:} We developed four deep learning-based pipelines for Type B aortic dissection segmentation: a single-step model, a sequential model, a sequential multi-task model, and an ensemble model, utilizing 3D U-Net and Swin-UnetR architectures. A dataset of 100 retrospective CTA images was split into training (n=80), validation (n=10), and testing (n=10). Performance was assessed using the Dice Coefficient and Hausdorff Distance.

    \noindent \textbf{Results:} Our approach achieved superior segmentation accuracy, with Dice Coefficients of 0.91 ± 0.07 for TL, 0.88 ± 0.18 for FL, and 0.47 ± 0.25 for FLT, outperforming Yao et al. \cite{yao_imagetbad_2021}, who reported 0.78 ± 0.20, 0.68 ± 0.18, and 0.25 ± 0.31, respectively.
    
    \noindent \textbf{Conclusion:} The proposed pipelines provide accurate segmentation of TBAD features, enabling derivation of morphological parameters for surveillance and treatment planning.
    
\end{abstract}




\end{frontmatter}




\section{Introduction}
\label{introduction}

Aortic dissection (AD) is a life-threatening cardiovascular condition involving a rupture in the aorta's inner layer, forming a true lumen (TL) and a false lumen (FL) \cite{10.1001/jama.283.7.897}. The FL can develop a blood clot, known as a false lumen thrombosis (FLT), which may narrow the TL and may increase rupture risk or be protective of rupture when fully thrombosed. AD is classified into Type A (TAAD), dissection of the ascending aorta plus or minus the descending aorta, and Type B (TBAD), dissection of the descending aorta only. While TAAD requires immediate surgery, TBAD can be managed either surgically or medically \cite{pmid25662791}.

\begin{figure*}[h]
    \centering
    \includegraphics[width=0.75\linewidth]{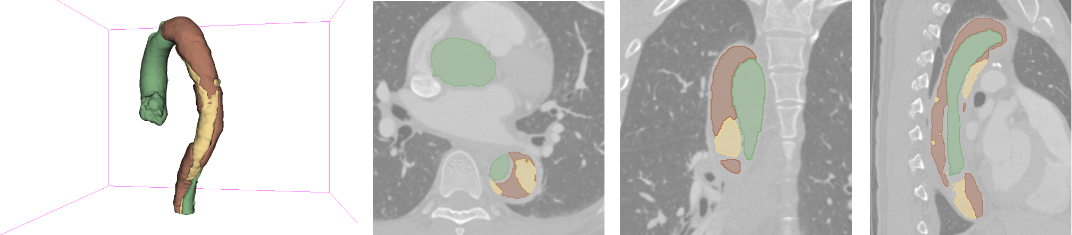}
    \caption{Visualization of a TBAD in a 3D model including FLT(Red), TL(Green), and FL(Yellow), and the corresponding CTA image with axial, coronal, and sagittal views}. 
    \label{fig:TBAD}
\end{figure*}

Computed Tomography Angiography (CTA) is critical for diagnosing and planning treatments for AD \cite{pmid27090166}, with accurate segmentation of TL, FL, and FLT being essential for prognosis and mortality risk assessment. Partial FL thrombosis accelerates aortic expansion compared to complete thrombosis or an unobstructed FL, requiring vigilant monitoring \cite{trimarchi_importance_2013}. Automated segmentation of the dissected aorta can assist in tracking changes over time, such as the calibre of the true lumen (TL) and false lumen (FL), which are critical for patient management. For instance, a progressively thrombosing false lumen may indicate stabilization, while a larger FL is a risk factor for rupture, and a smaller TL may increase the risk of end-organ ischemia. Manual segmentation, however, remains time-intensive, taking up to 1.5 hours per scan for radiologists \cite{yao_imagetbad_2021}, underscoring the need for automated solutions.

Deep learning (DL), particularly Convolutional Neural Networks (CNNs), has demonstrated promise in medical image segmentation by automating feature extraction \cite{ronneberger2015unet, 9098956, isensee_nnu-net_2021, pmid37328572}. Recently, transformer architectures \cite{dosovitskiy2021image, chen2021transunet, cao2021swinunet} and State Space Models (SSMs) like Mamba have further improved segmentation by capturing long-range dependencies \cite{gu2022efficiently, xing2024segmamba, ma2024umamba}.

Despite advancements, challenges persist in accurately segmenting TBAD features, particularly FLT \cite{yao_imagetbad_2021}. This study proposes DL-based pipelines to improve TL, FL, and FLT segmentation in TBAD CTA scans.

In summary, our main contributions are outlined below:
 \begin{itemize}
     \item \textbf{TBAD Segmentation Pipeline}: We developed an automatic segmentation pipeline that outperformed previous methods in segmenting TL, FL, and FLT from CTA scans.
     \item \textbf{Ensemble Approach for Improved Performance}:  Our ensemble model, combining outputs from two single-step models, showed superior performance, surpassing both previous methods and our other pipelines.
     \item \textbf{Analysis of Sequential Stage Models}: We explored the benefits and drawbacks of sequential models in segmentation tasks, highlighting their quicker convergence but potential dependency issues.
 \end{itemize}

\section{Materials and methods}

\subsection{Previous Studies}
Many studies have explored automatic segmentation utilizing deep learning. Cao et al. developed a sequential multi-stage CNN architecture for automated TBAD segmentation, using a 3D U-Net to roughly segment the aorta and generate a bounding box, followed by a second 3D U-Net to distinguish between the TL and FL based on the down-sampled input and bounding box \cite{pmid31683252}. Although down-sampling aids in overcoming hardware limitations, it leads to a quality loss in both CT images and ground truth labels \cite{9631067}.
Yao et al. introduced models integrating aorta segmentation with TL, FL, and FLT into unified outputs. While achieving comparable results for TL and FL, the FLT segmentation Dice coefficient (DC) was relatively low at around 0.25, indicating significant room for improvement \cite{yao_imagetbad_2021}. The use of down-sampling during preprocessing resulted in the loss of crucial information, especially when FLT comprised only a small portion of the aorta, and the lack of held-out cross-validation made test results less reliable.
Chen et al. used a multi-stage learning approach, incorporating a preprocessing method to straighten the aorta to mitigate curvature effects and simplify segmentation. They cropped image patches instead of down-sampling, achieving better results without significant information loss \cite{chen_multi-stage_2021}.
Wobben et al. developed three segmentation models using a 3D residual U-Net. The second model, incorporating a cascaded network for FLT segmentation, demonstrated the highest performance \cite{9631067}. However, prior knowledge of FLT presence is required for effective segmentation in this approach.
Xiang et al. enhanced this approach by integrating a flap attention network in their model to target the intimal flap, achieving 0.91 DC for TL and 0.88 DC for FL \cite{xiang_adseg_2023}. Although they included FLT cases, no specific FLT segmentation results were presented.
Hahn et al. employed 2D models using CTA slices and generating multi-planar reformatted images orthogonal to the aortic centerline for final segmentation outcomes \cite{hahn_ct-based_2020}. Despite promising results with this 2D approach, including a third dimension is crucial to differentiate between thrombus and slow blood flow for for accurate FLT segmentation.
These studies highlight the ongoing challenges and need for innovations in TBAD segmentation, particularly for accurately segmenting FLT.

\subsection{Imaging Data}
The ImageTBAD dataset consists of 100 retrospectively collected 3D CTA images from Guangdong Provincial People's Hospital, acquired from January 1, 2013, to April 23, 2015 \cite{yao_imagetbad_2021}. Images were acquired from two kinds of scanners (Siemens SOMATOM Force, and Philips 256-slice Brilliance iCT system). This dataset includes pre-operative CTA scans of TBAD, each sized 512 x 512 x (135-416), The images are segmented into TL, FL, and FLT by two expert cardiovascular radiologists. Out of the total cases, 68 have FLT, while 32 do not. All CTA cases are provided in Nifti format.
We implemented cross-validation with 5 folds to maximize the utilization of this small dataset. To ensure consistent model performance across all folds, we maintained a constant ratio of FLT cases within each fold, preserving the balance of the data. Preprocessing involved restricting the intensity range to -500 to 1000 Hounsfield Units (HU) to capture relevant anatomical structures. Each image was resampled to a uniform voxel size of 1.5 mm in all dimensions and cropped to focus on the foreground, removing irrelevant background. This procedure reduces data size and computational load. Data augmentation, including cropping, flipping, rotating, and intensity shifting, was applied to 20\% of the training data to enhance model robustness.

\subsection{Segmentation Pipeline}
\begin{figure*}[h]
\centering
\includegraphics[width=1\textwidth]{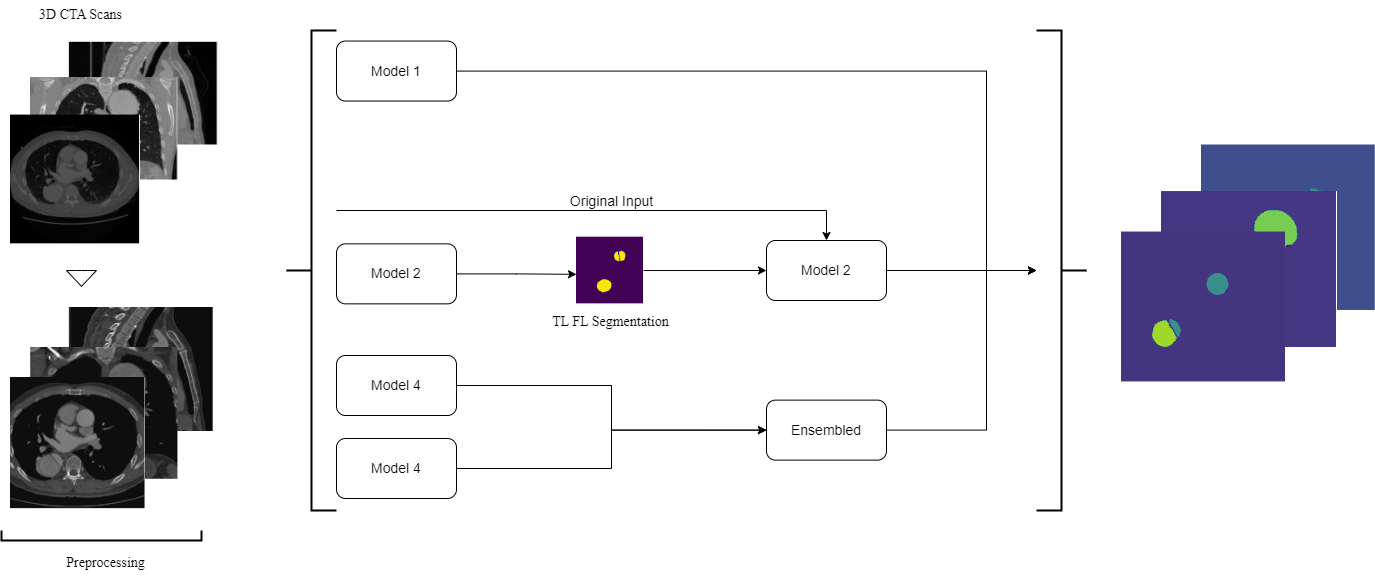}
\caption{Segmentation pipeline for Models 1, 2, and 4.}
\label{fig:124}
\end{figure*}

\begin{figure*}[h]
\centering
\includegraphics[width=1\textwidth]{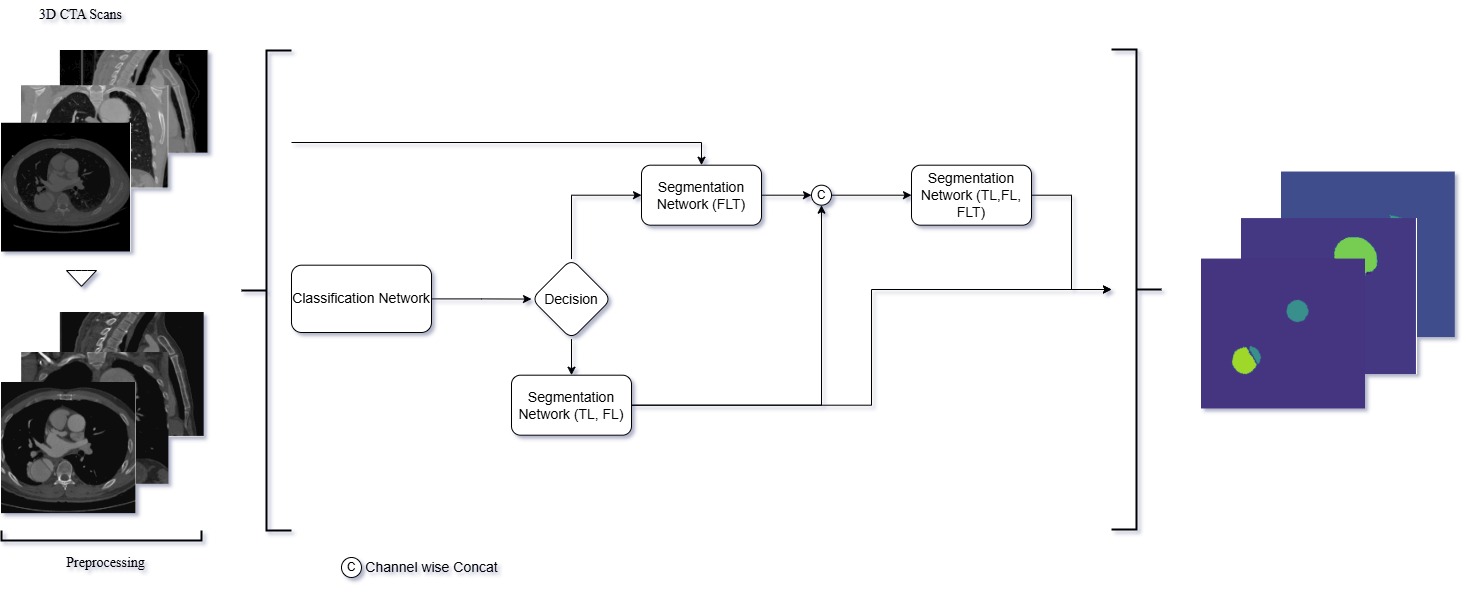}
\caption{Segmentation pipeline for Model 3.}
\label{fig:3}
\end{figure*}

We designed four segmentation pipelines: (1) a single-step model, (2) a sequential model, (3) a sequential multi-task model, and (4) an ensemble model. Pipelines 1, 2, and 4 are illustrated in Figure \ref{fig:124}, while Pipeline 3 is shown in Figure \ref{fig:3}. All pipelines take pre-processed CTA images as input and output the segmentation results of all classes.

The single-step model pipeline (Model 1) uses a single DL network to classify image voxels as TL, FL, FLT, and background, streamlining the segmentation process and enhancing computational efficiency. 
The sequential model pipeline (Model 2) employs a cascade of networks, with the first network segmenting the entire aorta and the second network segmenting TL, FL, and FLT using the original input and the first network's output as a mask. 
In the sequential multi-task model pipeline (Model 3), a classification model first determines the presence of FLT. If FLT is detected, the CTA image is processed by a segmentation network specifically trained to segment FLT. Simultaneously, another segmentation network focuses on segmenting the TL and FL. Finally, the outputs from the FLT segmentation and the TL/FL segmentation are concatenated with the original image and fed into a final network that produces the combined segmentation output. If FLT is not detected, the process continues without the FLT segmentation network. The major difference compared to Wobben et al.'s \cite{9631067} approach lies in the structure of the cascaded model. In their method, they first segmented the whole aorta and then combined this segmentation with the original input for the second network to segment the TL and FL. Finally, they combined all previous outputs to produce the final result. In contrast, our approach includes a dedicated FLT segmentation network, allowing for more targeted and accurate segmentation of FLT.
Finally, the ensemble model pipeline (Model 4) combines the outputs of SwinUNETR and 3D UNet by averaging their probabilities to improve segmentation accuracy.

We employed two segmentation networks, 3D U-Net \cite{çiçek20163d} and Swin-UnetR \cite{hatamizadeh2022swin}, and two classification networks, DenseNet-121 and DenseNet-264 \cite{huang2018densely}, selecting them based on their testing performance. For loss functions, we evaluated Generalized Dice Loss (GDL) \cite{Sudre_2017} and Dice Cross Entropy Loss (DCEL). GDL is suited for unbalanced datasets, while DCEL combines Dice Loss and Cross Entropy Loss to enhance performance. Based on superior validation results, DCEL was chosen as the primary loss function.

All methods were implemented using the PyTorch framework \cite{paszke2019pytorch}. Networks were trained on pre-processed and augmented data, with a batch size set to 1. A step learning rate approach was used, starting at \(1e^{-4}\) and reduced by a factor of 0.1 every 30 epochs. The Adam optimizer \cite{kingma_adam_2017} with a weight decay of \(1e^{-5}\) was employed. Training was conducted for a total of 50 epochs for the initial models in each fold, followed by an additional 20 epochs for the subsequent cascaded models that utilized the output of the earlier models. All training was performed on an Nvidia RTX 4070 Ti Super with 16 GB of VRAM.
We used the DC and Hausdorff Distance (HD) as evaluation metrics. The best-performing model was selected based on validation DC and True FLT DC. True FLT DC refers to the DC calculated specifically for cases that contain FLT, ensuring that the model's performance on FLT segmentation is accurately assessed. The original FLT DC, on the other hand, measures all cases, including those without FLT; in such cases, if the model incorrectly segments FLT where it doesn't exist, the DC will be 0. True FLT DC excludes cases where FLT is not present, providing a focused evaluation of FLT segmentation performance.
Classification model performance was evaluated using Precision, Recall, and F1 Score.

\section{Results}

In this section, we present the results of all our proposed models and methods. In Section \ref{segmentation_results}, we focus on the evaluation metrics, including the DC and HD for all proposed models, and also the performance of the classification model. To compare them, we have implemented the approach of Yao et al. \cite{yao_imagetbad_2021} on our dataset also by removing some suboptimal preprocessing steps of theirs which is discussed in Section \ref{Discussion}. In Section \ref{loss functions result}, we compare model performance and training details using different loss functions. Finally, in Section \ref{visualization}, we visualize the segmentation results of the models and compare them with the original ground truth.

\subsection{Segmentation}\label{segmentation_results}
In our evaluation of the segmentation performance, we compared the proposed methods against the baseline established by Yao et al \cite{yao_imagetbad_2021}. The primary metrics used were the DC and HD, assessing the accuracy and spatial precision of our models.

As shown in Table \ref{tab:model_evaluation_dc}, Method 3, the multi-task sequential model, achieved the highest DC across both FLT and True FLT, indicating superior segmentation performance, particularly in identifying and segmenting FLT. The performance on TL and FL was consistent across all models, with no significant differences observed. It is also worth noting that the classification network did not perform well in identifying FLT cases, with both classification networks achieving only around 50\% accuracy. As a result, the classification step was bypassed in Method 3. The True FLT metric, calculated on the testing set, further underscores the model's ability to accurately segment FLT when present.

\begin{table*}[ht]
    \centering
    \begin{tabularx}{\textwidth}{@{} l c *{4}{>{\centering\arraybackslash}X} @{}}
        \toprule
        \textbf{Method} & \textbf{Phase} & \textbf{TL} & \textbf{FL} & \textbf{FLT} & \textbf{True FLT} \\
        \midrule
        \multirow{1}{*}{Yao et al.'s} & Testing & 0.78 ± 0.20 & 0.68 ± 0.18 & 0.15 ± 0.32 & 0.25 ± 0.31 \\
        \midrule                        
        \multirow{2}{*}{Method 1} & Validation & 0.92 ± 0.04 & 0.87 ± 0.10 & 0.24 ± 0.25 & 0.40 ± 0.19 \\
                                  & Testing    & 0.92 ± 0.06 & 0.87 ± 0.08 & 0.36 ± 0.36 & 0.43 ± 0.26 \\
        \midrule
        \multirow{2}{*}{Method 2} & Validation & 0.93 ± 0.03 & 0.90 ± 0.08 & 0.21 ± 0.24 & 0.35 ±0.20 \\
                                  & Testing    & 0.92 ± 0.06 & \textbf{0.89 ± 0.08} & 0.19 ± 0.27 & 0.32 ± 0.29 \\
        \midrule
        \multirow{2}{*}{Method 3} & Validation & 0.92 ± 0.03 & 0.88 ± 0.09 & 0.26 ± 0.29 & 0.43 ± 0.26 \\
                                  & Testing    & 0.91 ± 0.07 & 0.88 ± 0.08 & \textbf{0.38 ± 0.36} & \textbf{0.47 ± 0.25} \\
        \midrule
        \multirow{2}{*}{Method 4} & Validation & 0.92 ± 0.04 & 0.86 ± 0.13 & 0.27 ± 0.28 & 0.45 ± 0.21 \\ & Testing    &\textbf{ 0.92 ± 0.05} & 0.88 ± 0.07 & 0.35 ± 0.38 & 0.42 ± 0.31 \\
        \bottomrule
    \end{tabularx}
    \caption{Models were validated and tested using the DC, including methods from Yao et al.  All reported values are the mean values across the entire validation/test set, with standard deviations indicated by ±. }
    \label{tab:model_evaluation_dc}
\end{table*}

Table \ref{tab:model_evaluation_hd} presents the HD results. In this table, FLT is only included in cases where it is present, as HD cannot be calculated when FLT is absent. Method 4 showed a significant advantage over the baseline and other methods in TL and FL segmentation, demonstrating the benefits of the ensemble approach. The results highlight the strengths and weaknesses of each method, with Method 4 providing the most balanced performance across all metrics.
\begin{table*}[ht]
    \centering
    \begin{tabularx}{\textwidth}{@{} l c *{3}{>{\centering\arraybackslash}X} @{}}
        \toprule
        \textbf{Phase} & \textbf{Method} & \textbf{TL} & \textbf{FL} & \textbf{FLT}\\
        \midrule
        \multirow{5}{*}{Testing} & Yao et al.'s & 96.87 ± 38.66 & 34.19 ± 25.81  & 93.41 ± 64.93\\
        & Method 1 & 65.56 ± 65.38 & 49.08 ± 71.20  & 76.71 ± 60.06\\
        & Method 2 & 67.77± 63.25 & 105.42 ± 79.85 & 110.57 ± 70.24\\
        & Method 3 & 76.72 ± 60.45 & 55.28 ± 43.83 & 81.97 ± 55.06\\
        & Method 4 & \textbf{24.05 ± 16.30} & \textbf{21.75 ± 17.51} & \textbf{74.25 ± 61.20}\\
        \bottomrule
    \end{tabularx}
    \caption{Models were tested using the HD, including methods from Yao et al. All reported values except FLT are the mean values across the entire test set, FLT HD is only calculated when FLT is presented in the case with standard deviations indicated by ±.}
    \label{tab:model_evaluation_hd}
\end{table*}

\subsection{Loss Functions} \label{loss functions result}
\begin{figure*}[h]
    \centering
    \begin{subfigure}[b]{0.45\linewidth}
        \includegraphics[width=\linewidth]{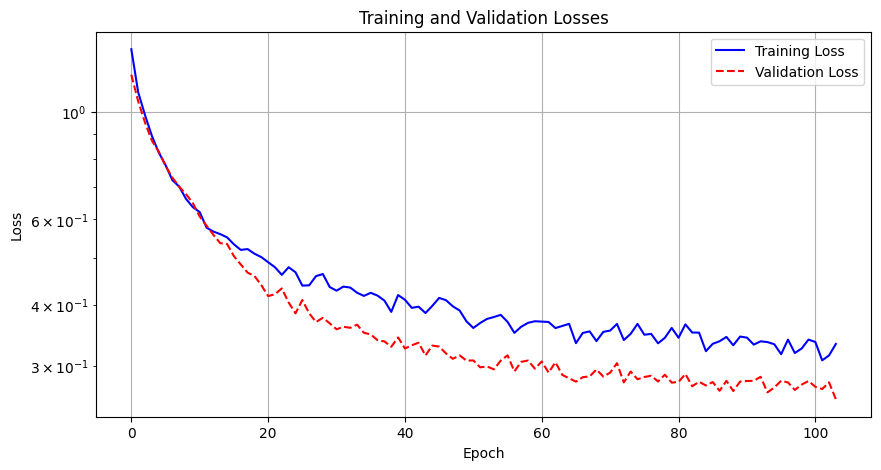}
        \caption{Training Swin-UnetR using DCEL}
        \label{fig:dce-loss}
    \end{subfigure}
    \hfill 
    \begin{subfigure}[b]{0.45\linewidth}
        \includegraphics[width=\linewidth]{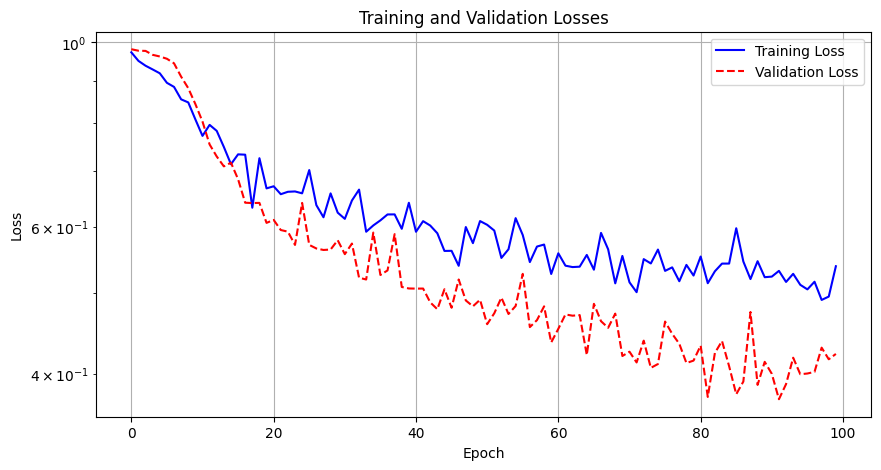}
        \caption{Training Swin-UnetR using GDL}
        \label{fig:gdl-loss}
    \end{subfigure}
    \caption{Training history comparison using DCEL and GDL on Swin-UnetR}
    \label{fig:training_process}
\end{figure*}

The impact of different loss functions on model performance was evaluated and presented in Table \ref{tab:loss_function_evaluation}. The results indicate that the Dice Cross Entropy Loss (DCEL) outperformed the Generalized Dice Loss (GDL) for Swin-UnetR across all classes. DCEL showed particularly strong performance in segmenting TL and FL. The training process, illustrated in Figure \ref{fig:training_process}, shows that models trained with DCEL converged more effectively compared to those trained with GDL.
\begin{table*}[ht]
    \centering
    \begin{tabularx}{\textwidth}{@{} l c *{4}{>{\centering\arraybackslash}X} @{}}
        \toprule
        \textbf{Network} & \textbf{Phase} & \textbf{TL} & \textbf{FL} & \textbf{FLT} & \textbf{True FLT}\\
        \midrule
        \multirow{2}{*}{3D U-Net} & GDL & 0.59 ± 0.22 & 0.72 ± 0.14 & \textbf{0.32 ± 0.36} & \textbf{0.37 ± 0.28}\\
                                 & DCEL & \textbf{0.78 ± 0.23} & \textbf{0.73 ± 0.18} & 0.22 ± 0.31 & 0.36 ± 0.32\\
        \midrule
        \multirow{2}{*}{Swin-UnetR} & GDL & 0.74 ± 0.09 & 0.74 ± 0.14 & 0.30 ± 0.36 & 0.34 ± 0.27\\
                                 & DCEL & \textbf{0.92 ± 0.06} & \textbf{0.87 ± 0.08} & \textbf{0.36 ± 0.36} & \textbf{0.43 ± 0.26} \\
        \bottomrule
    \end{tabularx}
    \caption{Loss function evaluation using DC with standard deviations indicated by ±.}
    \label{tab:loss_function_evaluation}
\end{table*}

\subsection{Visualization} \label{visualization}
\begin{figure*}[h]
    \centering
    \includegraphics[width=1\linewidth]{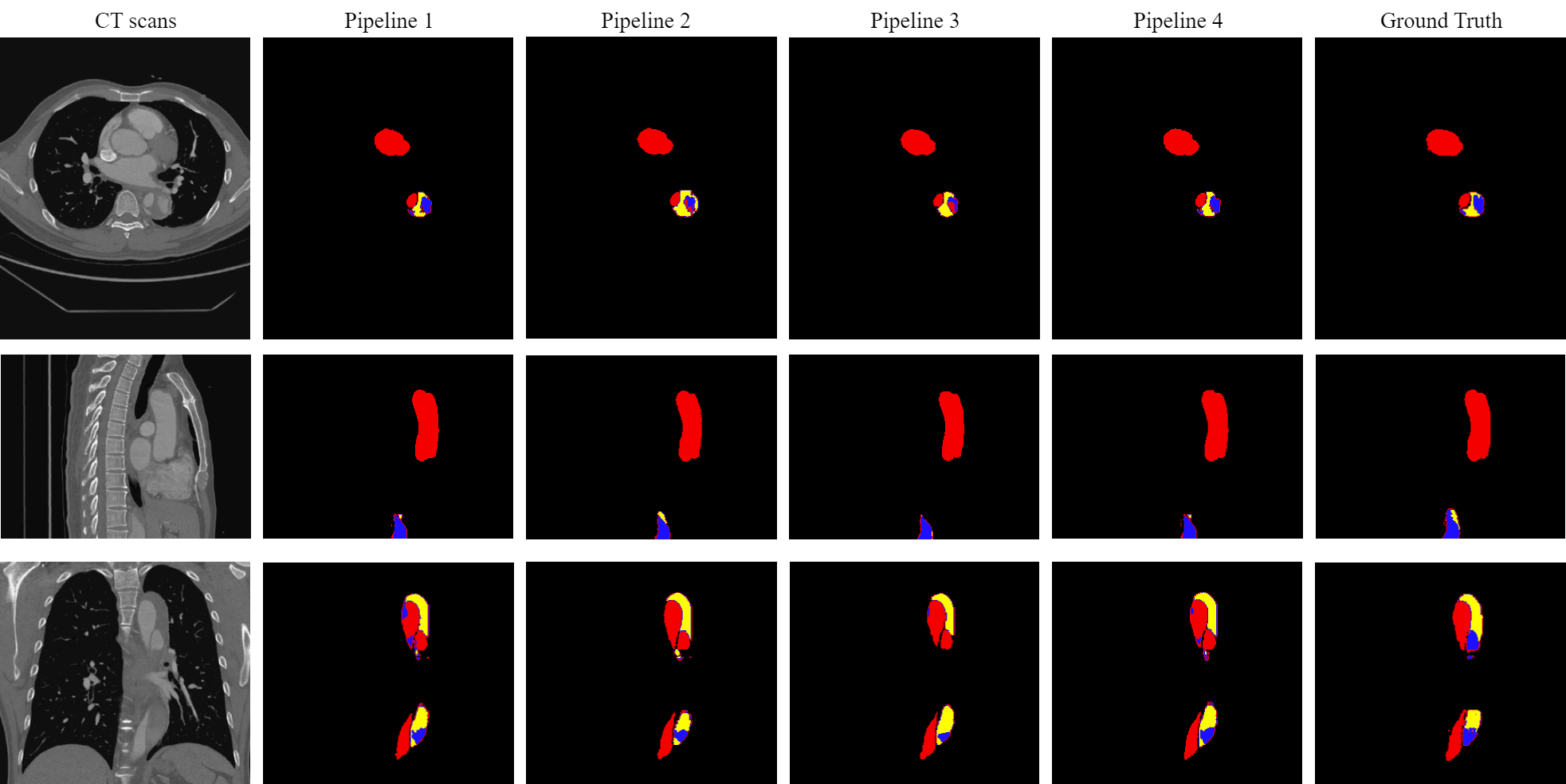}
    \caption{Segmentation result of all proposed models. }
    \label{fig:test-output}
\end{figure*}

We visualized the segmentation results of all proposed models and compared them with the original ground truth. Figure \ref{fig:test-output} showcases the performance of each model, highlighting their ability to accurately segment TL, FL, and FLT. Method 4 demonstrated the most precise and consistent segmentation results, aligning closely with the ground truth.

\section{Discussion} \label{Discussion}
Our study demonstrates significant improvements in TBAD segmentation, particularly in FLT segmentation, underscoring the efficacy of various deep learning models and methodologies. Notably, newer models like Swin-UnetR, which utilize transformer architecture, outperformed the classical 3D U-Net in organ-level segmentation. This suggests that transformers excel in tasks where capturing global context is crucial.

A cascade network with additional input information enables models with more parameters to learn better segmentation, as demonstrated by Methods 2 and 3. Method 3, which includes an additional network specifically focused on FLT segmentation, shows a significant performance improvement compared to Method 2, with the True FLT Dice Coefficient increasing from 0.32 to 0.47. However, it is also important to note that training a cascaded network significantly increases time consumption due to the larger data input and the time required for the base models to segment the inputs.

Combining results from different models significantly enhanced overall performance. The ensemble approach leveraged the strengths of each model, leading to more robust and accurate segmentation results. Interestingly, our findings indicate that simpler pipelines can yield better performance than more complex methodologies, emphasizing the importance of evaluating complexity against actual performance gains.

Our results surpass the previous work by Yao et al. \cite{yao_imagetbad_2021} by a significant margin, particularly in FLT segmentation. Notably, when we retrained Yao et al.'s model under the same conditions as our study, Method 3 achieved an average DC of 0.47 ± 0.25 for True FLT and Method 4 achieved 74.25 ± 61.20 for FLT in terms of HD, compared to their original method's performance of 0.25 ± 0.31 DC for True FLT and 93.41 ± 64.93 for HD. These results demonstrate the efficacy of our proposed methods. Furthermore, during the implementation of Yao et al.'s method, it was observed that their downsampling approach caused a significant loss of information, reducing the number of FLT cases from 68 to 60. This reduction likely contributed to the inaccuracies in their study's results and explains the substantial drop in model performance observed during our implementation after removing downsampling in preprocess.

Despite these advances, our study faced several limitations. The primary limitation was the scarcity of data, with the dataset containing only 100 cases and just 68 cases featuring FLT. This limited size is insufficient for DL models to achieve optimal parameter adjustment, potentially leading to over-fitting and reduced generalization to new data. Additionally, the quality of the dataset posed challenges; some cases were mislabeled, indicating FLT when none was present, which misled the DL model and affected its accuracy.

Another significant limitation was hardware constraints. 3D medical imaging segmentation requires substantial memory, and the large size of CTA images necessitated resampling cropping, and patch-based processing. These preprocessing steps can lead to the loss of important features and contextual information, impacting the model’s performance. High-performance GPUs with large memory capacities are essential, yet even with such hardware, training times remain long and computationally intensive.

Future work should focus on exploring more powerful DL networks, like the Mamba architecture, which may offer significant performance improvements. Leveraging pre-trained backbone networks specifically designed for 3D medical imaging tasks could also yield better performance for TBAD segmentation.

In conclusion, our study presents a DL approach to TBAD segmentation, demonstrating substantial improvements over existing methods. By integrating both classical and advanced DL models, and utilizing ensemble techniques, we achieved enhanced segmentation accuracy and robustness. Addressing the limitations related to data scarcity, quality, and hardware constraints is crucial for further advancements. Future research should explore more sophisticated models and validation techniques to continue improving TBAD segmentation and ultimately support better diagnosis and treatment planning for patients.



\bibliographystyle{vancouver} 
\bibliography{sample}

\begin{thebibliography}{10}

\bibitem{yao_imagetbad_2021}
Yao Z, Xie W, Zhang J, Dong Y, Qiu H, Yuan H, et~al.
\newblock {ImageTBAD}: A 3D Computed Tomography Angiography Image Dataset for Automatic Segmentation of Type-B Aortic Dissection.
\newblock Frontiers in Physiology. 2021;12.
\newblock Available from: \url{https://www.frontiersin.org/journals/physiology/articles/10.3389/fphys.2021.732711}.

\bibitem{10.1001/jama.283.7.897}
Hagan PG, Nienaber CA, Isselbacher EM, Bruckman D, Karavite DJ, Russman PL, et~al.
\newblock {The International Registry of Acute Aortic Dissection (IRAD)New Insights Into an Old Disease}.
\newblock JAMA. 2000 02;283(7):897-903.
\newblock Available from: \url{https://doi.org/10.1001/jama.283.7.897}.

\bibitem{pmid25662791}
Nienaber CA, Clough RE.
\newblock {{M}anagement of acute aortic dissection}.
\newblock Lancet. 2015 Feb;385(9970):800-11.

\bibitem{pmid27090166}
Kamman AV, van Herwaarden JA, Orrico M, Nauta FJ, Heijmen RH, Moll FL, et~al.
\newblock {{S}tandardized {P}rotocol to {A}nalyze {C}omputed {T}omography {I}maging of {T}ype {B} {A}ortic {D}issections}.
\newblock J Endovasc Ther. 2016 Jun;23(3):472-82.

\bibitem{trimarchi_importance_2013}
Trimarchi S, Tolenaar JL, Jonker FHW, Murray B, Tsai TT, Eagle KA, et~al.
\newblock Importance of false lumen thrombosis in type B aortic dissection prognosis.
\newblock The Journal of Thoracic and Cardiovascular Surgery. 2013;145(3):S208-12.
\newblock Available from: \url{https://www.sciencedirect.com/science/article/pii/S0022522312014894}.

\bibitem{ronneberger2015unet}
Ronneberger O, Fischer P, Brox T.
\newblock U-net: Convolutional networks for biomedical image segmentation.
\newblock In: Medical image computing and computer-assisted intervention--MICCAI 2015: 18th international conference, Munich, Germany, October 5-9, 2015, proceedings, part III 18. Springer; 2015. p. 234-41.

\bibitem{9098956}
Fan DP, Zhou T, Ji GP, Zhou Y, Chen G, Fu H, et~al.
\newblock Inf-Net: Automatic COVID-19 Lung Infection Segmentation From CT Images.
\newblock IEEE Transactions on Medical Imaging. 2020;39(8):2626-37.

\bibitem{isensee_nnu-net_2021}
Isensee F, Jaeger PF, Kohl SAA, Petersen J, Maier-Hein KH.
\newblock {nnU}-Net: a self-configuring method for deep learning-based biomedical image segmentation.
\newblock Nat Methods;18(2):203-11.
\newblock Publisher: Nature Publishing Group.
\newblock Available from: \url{https://www.nature.com/articles/s41592-020-01008-z}.

\bibitem{pmid37328572}
Dumitru RG, Peteleaza D, Craciun C.
\newblock {{U}sing {D}{U}{C}{K}-{N}et for polyp image segmentation}.
\newblock Sci Rep. 2023 Jun;13(1):9803.

\bibitem{dosovitskiy2021image}
Dosovitskiy A, Beyer L, Kolesnikov A, Weissenborn D, Zhai X, Unterthiner T, et~al.. An Image is Worth 16x16 Words: Transformers for Image Recognition at Scale; 2021.

\bibitem{chen2021transunet}
Chen J, Lu Y, Yu Q, Luo X, Adeli E, Wang Y, et~al.. TransUNet: Transformers Make Strong Encoders for Medical Image Segmentation; 2021.

\bibitem{cao2021swinunet}
Cao H, Wang Y, Chen J, Jiang D, Zhang X, Tian Q, et~al.. Swin-Unet: Unet-like Pure Transformer for Medical Image Segmentation; 2021.

\bibitem{gu2022efficiently}
Gu A, Goel K, Ré C. Efficiently Modeling Long Sequences with Structured State Spaces; 2022.

\bibitem{xing2024segmamba}
Xing Z, Ye T, Yang Y, Liu G, Zhu L. SegMamba: Long-range Sequential Modeling Mamba For 3D Medical Image Segmentation; 2024.

\bibitem{ma2024umamba}
Ma J, Li F, Wang B. U-Mamba: Enhancing Long-range Dependency for Biomedical Image Segmentation; 2024.

\bibitem{pmid31683252}
Cao L, Shi R, Ge Y, Xing L, Zuo P, Jia Y, et~al.
\newblock {{F}ully automatic segmentation of type {B} aortic dissection from {C}{T}{A} images enabled by deep learning}.
\newblock Eur J Radiol. 2019 Dec;121:108713.

\bibitem{9631067}
Wobben LD, Codari M, Mistelbauer G, Pepe A, Higashigaito K, Hahn LD, et~al.
\newblock Deep Learning-Based 3D Segmentation of True Lumen, False Lumen, and False Lumen Thrombosis in Type-B Aortic Dissection.
\newblock In: 2021 43rd Annual International Conference of the IEEE Engineering in Medicine \& Biology Society (EMBC); 2021. p. 3912-5.

\bibitem{chen_multi-stage_2021}
Chen D, Zhang X, Mei Y, Liao F, Xu H, Li Z, et~al.
\newblock Multi-stage learning for segmentation of aortic dissections using a prior aortic anatomy simplification.
\newblock Medical Image Analysis. 2021;69:101931.
\newblock Available from: \url{https://www.sciencedirect.com/science/article/pii/S1361841520302954}.

\bibitem{xiang_adseg_2023}
Xiang D, Qi J, Wen Y, Zhao H, Zhang X, Qin J, et~al.
\newblock {ADSeg}: A flap-attention-based deep learning approach for aortic dissection segmentation.
\newblock {PATTER}. 2023;4(5).
\newblock Publisher: Elsevier.
\newblock Available from: \url{https://www.cell.com/patterns/abstract/S2666-3899(23)00067-3}.

\bibitem{hahn_ct-based_2020}
Hahn LD, Mistelbauer G, Higashigaito K, Koci M, Willemink MJ, Sailer AM, et~al.
\newblock {CT}-based True- and False-Lumen Segmentation in Type B Aortic Dissection Using Machine Learning.
\newblock Radiol Cardiothorac Imaging. 2020;2(3):e190179.

\bibitem{çiçek20163d}
Özgün Çiçek, Abdulkadir A, Lienkamp SS, Brox T, Ronneberger O. 3D U-Net: Learning Dense Volumetric Segmentation from Sparse Annotation; 2016.

\bibitem{hatamizadeh2022swin}
Hatamizadeh A, Nath V, Tang Y, Yang D, Roth H, Xu D. Swin UNETR: Swin Transformers for Semantic Segmentation of Brain Tumors in MRI Images; 2022.

\bibitem{huang2018densely}
Huang G, Liu Z, van~der Maaten L, Weinberger KQ. Densely Connected Convolutional Networks; 2018.

\bibitem{Sudre_2017}
Sudre CH, Li W, Vercauteren T, Ourselin S, Jorge~Cardoso M.
\newblock In: Generalised Dice Overlap as a Deep Learning Loss Function for Highly Unbalanced Segmentations. Springer International Publishing; 2017. p. 240–248.
\newblock Available from: \url{http://dx.doi.org/10.1007/978-3-319-67558-9_28}.

\bibitem{paszke2019pytorch}
Paszke A, Gross S, Massa F, Lerer A, Bradbury J, Chanan G, et~al.. PyTorch: An Imperative Style, High-Performance Deep Learning Library; 2019.

\bibitem{kingma_adam_2017}
Kingma DP, Ba J. Adam: A Method for Stochastic Optimization. {arXiv};.
\newblock Available from: \url{http://arxiv.org/abs/1412.6980}.

\end{thebibliography}

\end{document}